\documentstyle[prl,aps,multicol,psfig]{revtex}
\begin{document}
\draft
\title{RESPONSE OF THE TWO-DIMENSIONAL ELECTRON GAS OF AlGaAs/GaAs
HETEROSTRUCTURES TO PARALLEL MAGNETIC FIELD}
\author{V.S.~Khrapai, E.V.~Deviatov, A.A.~Shashkin, V.T.~Dolgopolov}
\address{Institute of Solid State Physics, Chernogolovka, Moscow
District 142432, Russia}
\maketitle

\begin{abstract}
We study the transport properties of the two-dimensional electron gas
in AlGaAs/GaAs heterostructures in parallel to the interface magnetic
fields at low temperatures. The magnetoresistance in the metallic
phase is found to be positive and weakly anisotropic with respect to
the orientation of the in-plane magnetic field and the current
through the sample. At low electron densities ($n_s< 5\times
10^{10}$~cm$^{-2}$) the experimental data can be described adequately
within spin-related approach while at high $n_s$ the
magnetoresistance mechanism changes as inferred from
$n_s$-independence of the normalized magnetoresistance.
\end{abstract}
\begin{multicols}{2}

Much interest has been aroused recently by the behaviour of
two-dimensional (2D) electron systems in a parallel magnetic field.
The resistance of a 2D electron gas in Si MOSFETs was found to rise
steeply with parallel field $B_\parallel$ saturating to a constant
value above a critical magnetic field $B_c$ which depends on electron
density \cite{Sim,Krav,Mertes,Pud}. Such a behaviour of the
resistance agrees well with data on the metal-insulator phase diagram
for the case of parallel fields of Ref.~\cite{dol} where the
suppression of the metallic phase by $B_\parallel$ was observed. The
insensitivity of the effect to the orientation of $B_\parallel$ with
respect to the current through the sample \cite{Sim} as well as its
isotropy in weak, tilted magnetic fields \cite{Krav,dol} hinted at
the spin origin of the effect. Recently, an analysis of Shubnikov-de
Haas oscillations in tilted magnetic fields has established that the
field $B_c$ corresponds to the onset of full spin polarization of the
electron system \cite{Ok,Vit}. The influence of $B_\parallel$ on the
resistance of the 2D hole gas in GaAs heterostructures was found to
be basically similar to the case of Si MOSFETs \cite{sim} with two
noteworthy distinctions: (i) above $B_c$, the resistance keeps on
increasing less steeply with no sign of saturation \cite{Yoon}; and
(ii) the magnetoresistance is strongly anisotropic depending upon the
relative orientation of the in-plane magnetic field and the current
\cite{Pap}.

The early version of the theory of the spin origin of parallel field
effects exploits scaling arguments for calculating the
temperature-dependent magnetoresistance in the metallic phase in the
low field limit \cite{CC}. An alternative concept has been expressed
recently based on the fact that the 2D electron screening of a random
potential depends on the relative population of spin-up and spin-down
subbands \cite{Gold}: at zero temperature the magnetoresistance is
expected to be positive for relatively low electron densities in the
metallic phase and is determined by the spin polarization of 2D
electrons which is defined as the ratio of the Zeeman splitting and
the Fermi energy $\xi=g\mu_B B/2E_F$. Above the critical field $B_c$
corresponding to the condition $\xi=1$, the resistance
$R(B_\parallel)$ should saturate at the level of the four-fold
zero-field resistance. While the spin-related approach \cite{Gold}
allows the interpretation of the resistance rise with $B_\parallel$
in Si MOSFETs, the strongly anisotropic magnetoresistance observed on
the 2D holes in GaAs heterostructures is likely to point to a
contribution of the orbital effects of Ref.~\cite{Sarma} where it was
shown that for a 2D system with finite thickness the form of the
Fermi surface changes in a parallel magnetic field.

In a number of recent publications, the occurrence of a
zero-magnetic-field metal-insulator transition in Si MOSFETs and for
the 2D holes in GaAs as well as the origin of the effects observed in
parallel magnetic fields have been attributed to strong
particle-particle interaction as characterised by the Wigner-Seitz
radius $r_s$ (see, e.g., Ref.~\cite{Ok}). Oppositely, for the 2D
electrons in GaAs heterostructures the values of $r_s$ are almost an
order of magnitude lower, particularly, because of the small
effective mass, which is traditionally expressed in terms of weak
electron-electron interaction in GaAs. Nevertheless, for the 2D
electrons in both GaAs \cite{shash1} and Si MOSFETs \cite{shash},
similar metal-insulator phase diagrams were obtained in normal
magnetic fields including a zero field \cite{rem} and so the
parameter $r_s$ is not crucial in this case. The obvious consequence
of the small effective electron mass in GaAs is that much higher
values of the critical magnetic field $B_c$ are expected \cite{Gold}.

Here, we investigate the influence of parallel magnetic field on the
resistance in the metallic phase of the 2D electron system in GaAs
heterostructures. We observe a positive magnetoresistance which is
weakly anisotropic with respect to the orientation of the in-plane
magnetic field and the current through the sample. This finding is
similar to results reported for the 2D electrons in Si MOSFETs and
the 2D holes in GaAs heterostructures and enables us to split the
parallel magnetic field effect from the problem of the
interaction-induced metal-insulator transition. The spin mechanism
allows the description of the experimental data at low electron
densities but fails in the opposite limit in which the normalized
magnetoresistance is found to be independent of $n_s$.
\vbox{
\vspace{4mm}
\hbox{
\hspace{-0.2in}
\psfig{file=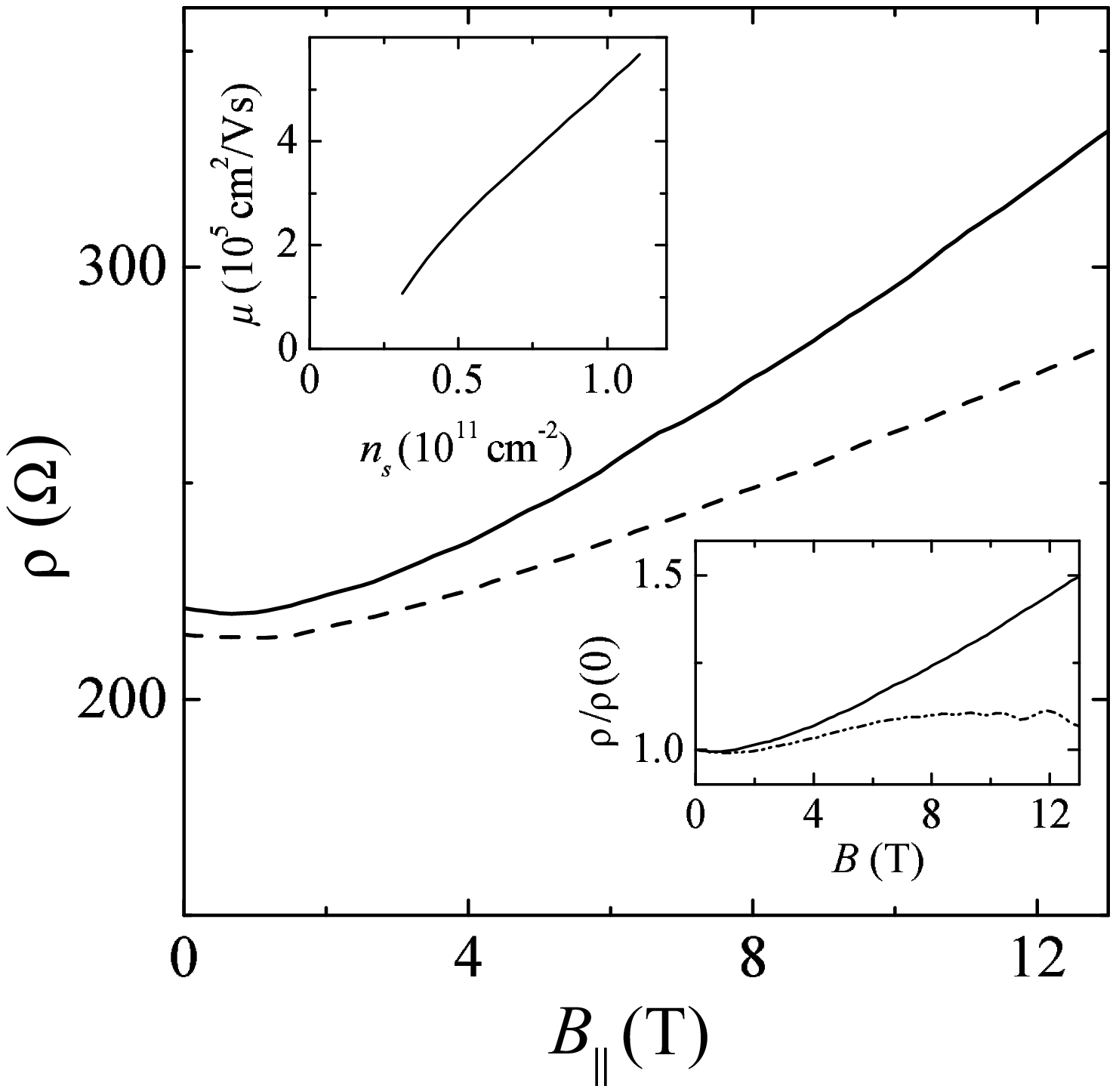,width=3.2in,bbllx=.5in,bblly=1.25in,bburx=7.25in,bbury=9.5in,angle=0}
}
\vspace{-1.2in}
\hbox{
\hspace{-0.15in}
\refstepcounter{figure}
\parbox[b]{3.4in}{\baselineskip=12pt \egtrm FIG.~\thefigure.
Dependence of the sample resistivity on parallel magnetic field for
$B_\parallel\perp I$ at $n_s=7.4\times 10^{10}$~cm$^{-2}$ (solid
line) and $B_\parallel\parallel I$ at $n_s=7.5\times
10^{10}$~cm$^{-2}$ (dashed line). The top inset shows the mobility as
a function of electron density. Bottom inset: behaviour of the
normalized resistivity with magnetic field in the perpendicular field
to current orientation for the in-plane $B$ (solid line) and the
field tilted by $1.2^\circ$ relative to the sample plane (dashed
line).\vspace{0.20in}
}
\label{f1}
}
}

Our devices are 170~$\mu$m wide conventional Hall bars based on an
AlGaAs/GaAs heterostructure that is grown on a (100) GaAs substrate
and contains a high mobility 2D electron gas. The density $n_s$ of
the 2D electrons is controlled using a gate on the front surface of
the device. The behaviour of the low-temperature mobility $\mu$ in
the studied range of electron densities is depicted in the top inset
to Fig.~\ref{f1}. For our samples the conductivity remains in the
metallic regime down to $n_s\sim 2\times 10^{10}$~cm$^{-2}$. The
sample is placed in the mixing chamber of a dilution refrigerator
with a base temperature of 30~mK. The measurements are performed
using a standard four-terminal lock-in technique at a frequency of
10~Hz in magnetic fields up to 14~T. The ac current $I$ through the
device does not exceed 1~nA. Two samples made from the same wafer
have been investigated; the results obtained on these are practically
identical.

To change the sample position in the mixing chamber we warm the
sample up, rotate it at room temperature, and cool it down again. The
alignment uncertainty of the sample plane with the magnetic field is
kept within $0.3^\circ$. We use small misalignments $\le 2^\circ$ to
observe quantum oscillations caused by a perpendicular component of
the magnetic field and evaluate the $g$ factor from the oscillation
beating pattern \cite{fang}, see the bottom inset to Fig.~\ref{f2}.
The electron density as a function of gate voltage is determined from
quantum oscillations in normal magnetic fields. We have checked
\vbox{
\vspace{7mm}
\hbox{
\hspace{-0.2in}
\psfig{file=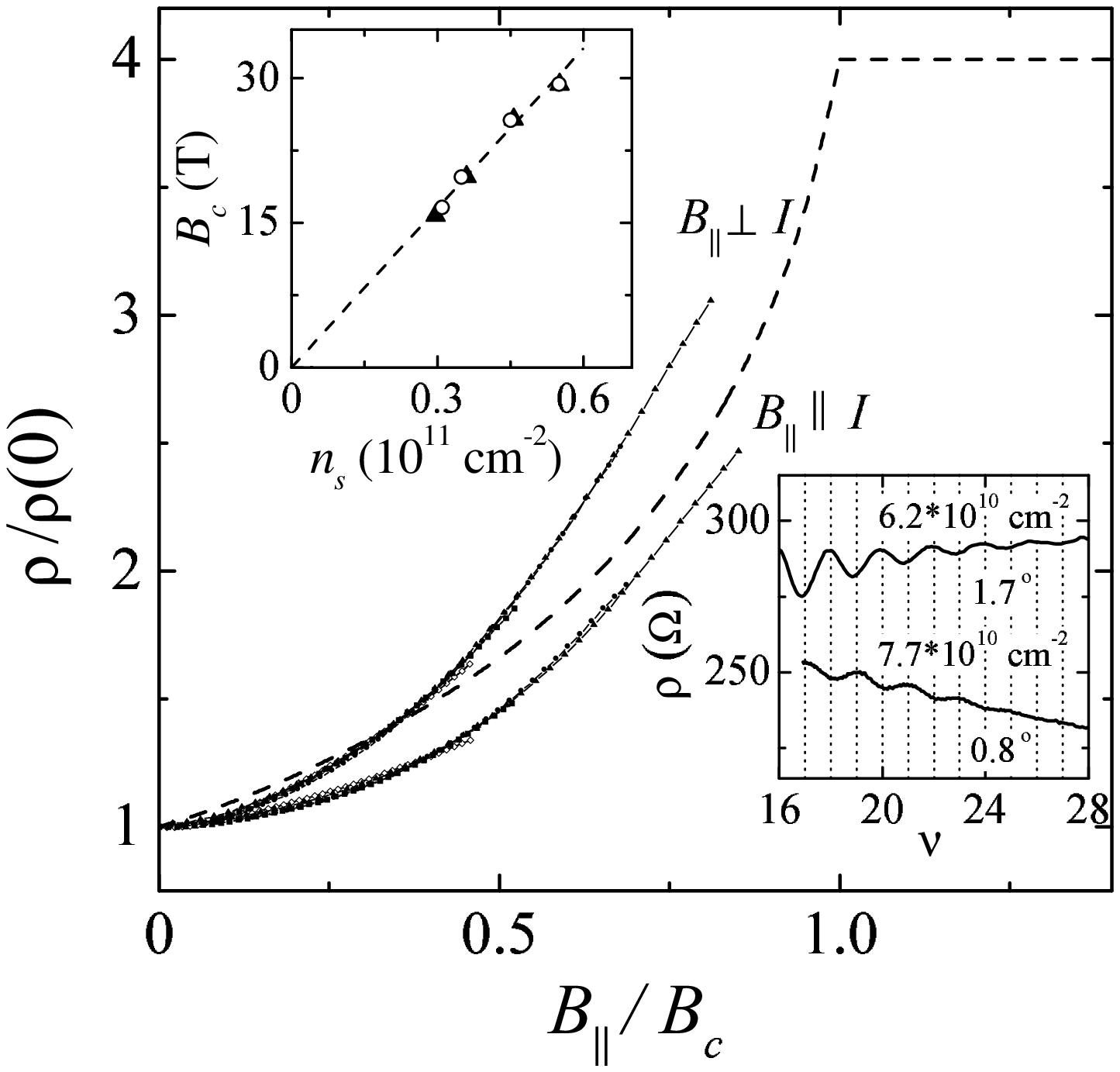,width=3.2in,bbllx=.5in,bblly=1.25in,bburx=7.25in,bbury=9.5in,angle=0}
}
\vspace{-1.4in}
\hbox{
\hspace{-0.15in}
\refstepcounter{figure}
\parbox[b]{3.4in}{\baselineskip=12pt \egtrm FIG.~\thefigure.
Scaling the magnetic field dependence of the normalized resistivity
of the sample at low $n_s$. Also shown by a dashed line is a fit
using the theory of Ref.~\protect\cite{Gold}. The scaling parameter
$B_c$ as a function of electron density is displayed in the top
inset. The bottom inset shows the resistivity as a function of
filling factor $\nu$ for two tilt angles of the magnetic field.
\vspace{0.20in}
}
\label{f2}
}
}
that
the gate voltage dependence of the resistance at $B=0$ is well
reproducible in different coolings of the sample with an accuracy of
insignificant threshold shifts. In parallel magnetic fields, this
dependence is used for determining the threshold voltage.

A typical experimental trace of the resistivity $\rho(B_\parallel)$
is shown in Fig.~\ref{f1} for the parallel and perpendicular
orientations of $B_\parallel$ relative to the current $I$. The
magnetoresistance is close to a parabolic dependence, being smaller
in the parallel configuration. As is evident from Fig.~\ref{f1}, the
magnetoresistance anisotropy is not strong, approximately a factor of
1.2. The observed resistance rise reaches a factor of three at the
lowest $n_s$ and the highest magnetic fields, displaying no tendency
of saturation. The temperature dependence of the resistance is
practically absent in the interval between 30 and 600~mK. We
emphasize that in the presence of a small normal component of the
magnetic field as caused by misalignment the dependence
$\rho(B_\parallel)$ is drastically distorted. As seen in the bottom
inset to Fig.~\ref{f1}, the effect is dramatic even for small
misalignments $\sim 1^\circ$.

By scaling the $B_\parallel$-axis we make the normalized resistivity
traces $\rho(B_\parallel)/\rho(0)$ at different electron densities
collapse onto a single curve simultaneously for each of the two field
directions (Fig.~\ref{f2}). At low $n_s$ the scaling factor $B_c$ is
found to enhance approximately linearly with electron density (top
inset to Fig.~\ref{f2}). At higher $n_s>5\times 10^{10}$~cm$^{-2}$
the normalized resistivity $\rho(B_\parallel)/\rho(0)$ becomes
independent of $n_s$ (Fig.~\ref{f3}) such that the scaling parameter
$B_c$ saturates.

Thus, we observe a strong rise of the resistance with
\vbox{
\vspace{5mm}
\hbox{
\hspace{-0.2in}
\psfig{file=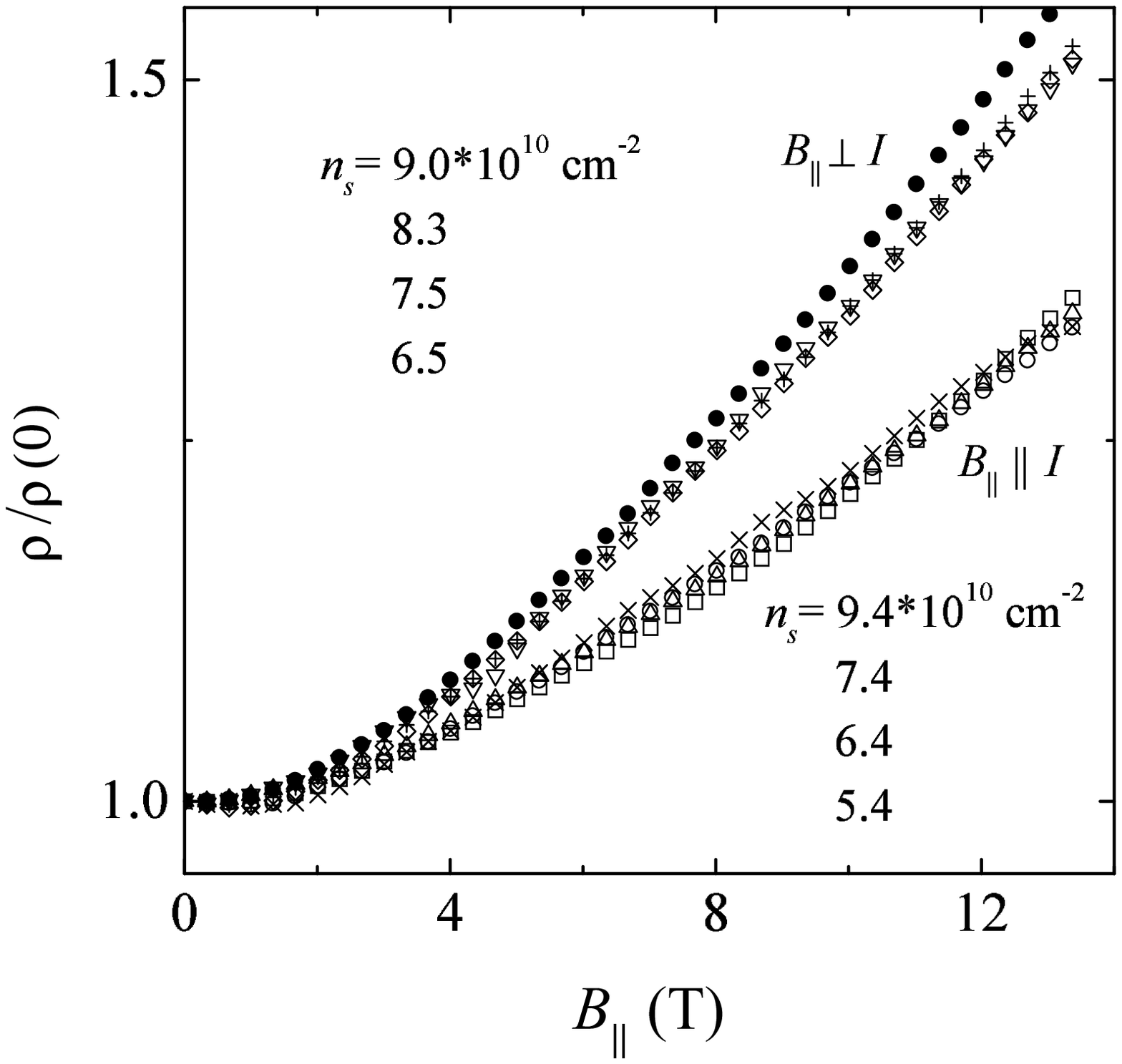,width=3.3in,bbllx=0in,bblly=1.25in,bburx=7.25in,bbury=9.5in,angle=0}
}
\vspace{-1.1in}
\hbox{
\hspace{-0.15in}
\refstepcounter{figure}
\parbox[b]{3.4in}{\baselineskip=12pt \egtrm FIG.~\thefigure.
Change of the normalized resistivity of the sample with parallel
magnetic field at high electron densities.\vspace{0.20in}
}
\label{f3}
}
}
parallel
magnetic field in the metallic regime in the 2D electron system with
$r_s$ spanning between 2 and 3.5, the range in which $\rho(0)$
changes by more than an order of magnitude. The effect is weakly
anisotropic relative to the orientation of $B_\parallel$ and $I$ and
is qualitatively similar to that found in 2D systems with strong
particle-particle interaction, $r_s\agt 10$
\cite{Sim,Krav,Mertes,Pud,sim,Yoon}. In contrast to the conclusion of
Refs.~\cite{Sim,Krav,Mertes,Pud,sim,Yoon}, we do not suppose that the
observed dependence $\rho(B_\parallel)$ is due to electron-electron
interaction because in our case the $r_s$ values are considerably
smaller. Particularly, the parallel field effect can be considered
regardless of the interaction-induced metal-insulator transition.

Two different approaches that predict the change of $\rho$ with
$B_\parallel$ in a 2D system with weak electron-electron interaction
have been formulated \cite{Gold,Sarma}. To compare our data with the
theoretical predictions we reason as follows. The absence of strong
anisotropy of the observed magnetoresistance allows one to presume
the dominance of spin effects as discussed in the theory of
Ref.~\cite{Gold}. This theory demands, in particular, identifying
$B_\parallel/B_c$ with the spin polarization $\xi$, i.e.,
$B_c=2E_F/g\mu_B$ (where $E_F=\pi \hbar^2n_s/m$ and $m$ is the
effective mass). The dashed line in Fig.~\ref{f2} is drawn in
accordance with the theory; its best fit to the data as shown in the
figure yields the normalizing condition for the parameter $B_c$.
Although the consistency between experiment and theory is fairly
good, there are problems with such a description of the data.
Firstly, the accessible magnetic fields are not high enough to reach
the expected saturation of the resistance. Secondly, the scaling
parameter $B_c$ becomes independent of electron density at
$n_s>5\times 10^{10}$~cm$^{-2}$. Thirdly, the so-defined critical
field $B_c(n_s)$ corresponds to the $g$ factor $g\approx 2.2$ which
is much larger than its bulk GaAs value of $g=0.44$. In fact, at low
electron
\vbox{
\vspace{4mm}
\hbox{
\hspace{-0.2in}
\psfig{file=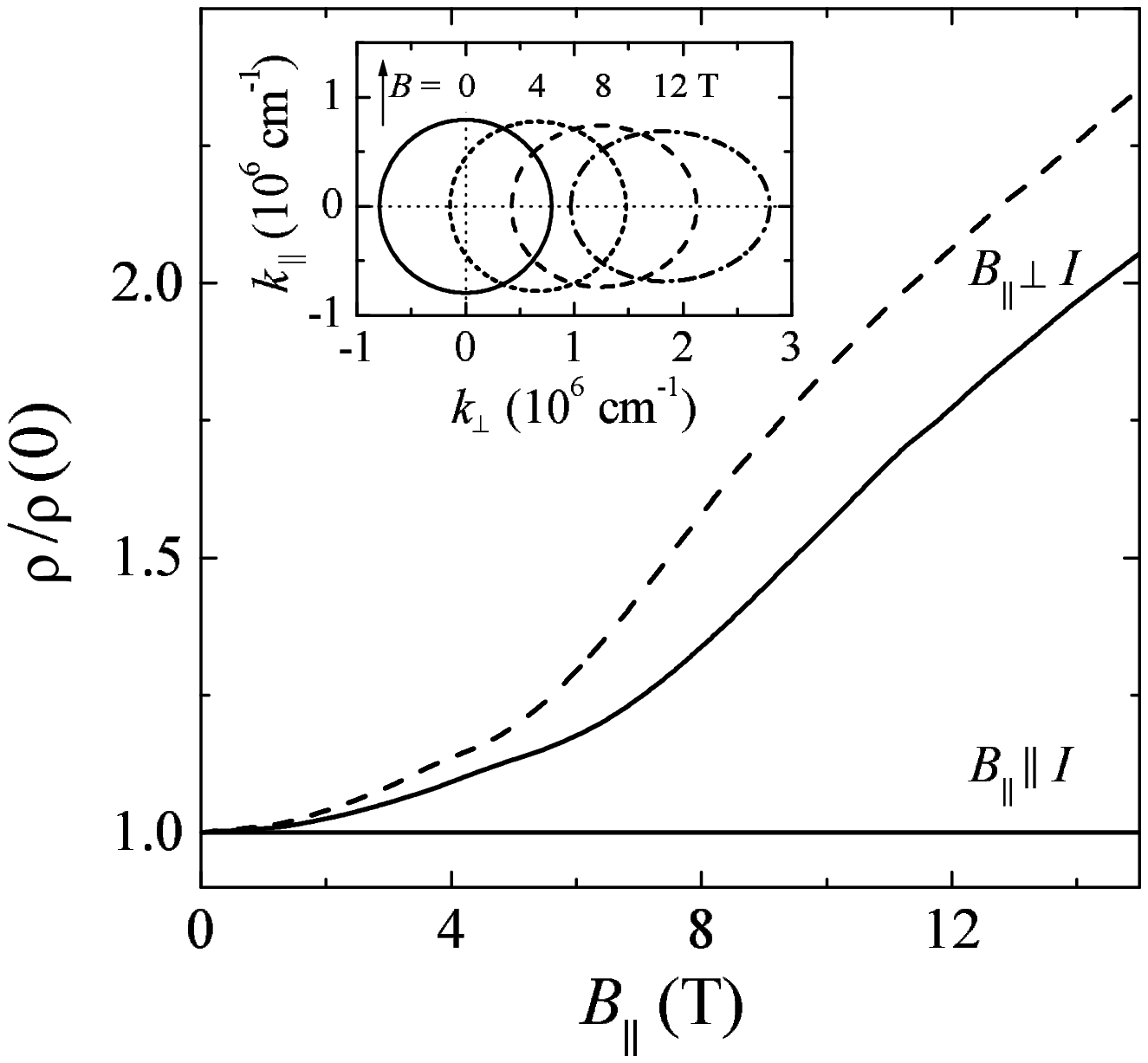,width=3.5in,bbllx=.5in,bblly=1.25in,bburx=7.25in,bbury=9.5in,angle=0}
}
\vspace{-1.8in}
\hbox{
\hspace{-0.15in}
\refstepcounter{figure}
\parbox[b]{3.4in}{\baselineskip=12pt \egtrm FIG.~\thefigure.
Magnetoresistance calculated in a similar way to the theory of
Ref.~\protect\cite{Sarma} for $n_s=1\times 10^{11}$~cm$^{-2}$ (solid
lines) and $n_s=7\times 10^{10}$~cm$^{-2}$ (dashed lines). The
corresponding transformation of the Fermi surface with parallel
magnetic field at $n_s=1\times 10^{11}$~cm$^{-2}$ is depicted in the
inset.\vspace{0.20in}
}
\label{f4}
}
}
densities the $g$ factor is expected to be enhanced due to
electron-electron interaction as discussed in the Fermi liquid model
(see, e.g., Ref.~\cite{ando}). An independent check of the beating
pattern of Shubnikov-de Haas oscillations in slightly nonparallel
magnetic fields gives a crude estimate $0.7 <g< 1.4$, which is still
small compared to the parallel field data \cite{note}. The noticeably
smaller values of $B_c$ obtained in the experiment as well as the
saturation of $B_c$ at high electron densities are in contrast to the
behaviour of the critical field $B_c$ found in Refs.~\cite{Ok,Yoon}
and cause us to invoke alternative mechanisms of the parallel field
magnetoresistance. The most likely candidate is an orbital effect
caused by the finite thickness of the 2D electron system
\cite{Sarma}. Its contribution would naturally explain the observed
magnetoresistance anisotropy and weaker dependence $B_c(n_s)$.
Nevertheless, we believe that at low electron densities the
spin-related concept \cite{Gold} describes the magnetoresistance
adequately because with decreasing $n_s$ the orbital effect (or any
other mechanism that yields $n_s$-independent scaling parameter
$B_c$) is overpowered by the spin effect as will be discussed below.

In a 2D electron system with finite thickness the parallel magnetic
field deforms the Fermi surface so that the effective mass in the
normal to $B_\parallel$ direction increases leading to a positive
magnetoresistance, whereas the one in the parallel direction remains
unchanged and so the resistance \cite{Sarma}. An example of the
deformed Fermi surface as calculated in triangular potential
approximation is displayed in the inset to Fig.~\ref{f4} for
different magnitudes of $B_\parallel$. With increasing magnetic field
the Fermi surface broadens in the $k_\perp$ direction and narrows in
the $k_\parallel$ direction to keep its area constant, shifting as a
whole along $k_\perp$. For lower $n_s$, the distortion of the Fermi
surface is stronger and so the magnetoresistance is larger, see
Fig.~\ref{f4}. Although the model \cite{Sarma} yields the correct
order of magnitude of the magnetoresistance for the perpendicular
field to current orientation (cf. Figs.~\ref{f3} and \ref{f4}), it
cannot describe the relatively weak magnetoresistance anisotropy as
well as $\rho(B_\parallel)/\rho(0)$ at different $n_s$. At the same
time, the theoretical magnetoresistance changes with electron density
not so strongly as the one predicted by the spin-related model
\cite{Gold}. Apparently, this is the condition for switching the
dominant magnetoresistance mechanism: at low $n_s$ the spin mechanism
of the magnetoresistance prevails while at high $n_s$ the orbital
effect is likely to become dominant.

In our opinion, the approach of Ref.~\cite{Sarma} should be completed
by including a change of the relaxation time in a parallel magnetic
field. The following aspects seem important: (i) the increase of the
Fermi surface perimeter leads to shortening the relaxation time; (ii)
the increase of the density of states at the Fermi energy results in
a better screening by the 2D system and, hence, increasing the
relaxation time; and (iii) the anisotropy of screening. Their account
should cause, at least, a reduction of the anisotropy of the
theoretical magnetoresistance.

In summary, we have investigated the transport properties of the 2D
electrons in AlGaAs/GaAs heterostructures in parallel to the
interface magnetic fields at low temperatures. It has been found that
the magnetoresistance in the metallic phase is positive and weakly
anisotropic with respect to the orientation of the in-plane magnetic
field and the current through the sample. This is basically similar
to data obtained for the 2D electrons in Si MOSFETs and the 2D holes
in GaAs heterostructures, although the electron-electron interaction
in GaAs is considerably weaker. Therefore, our experiment splits the
parallel magnetic field effect from the problem of the
interaction-induced metal-insulator transition. At low electron
densities the spin-related model is capable of describing the
experimental results while at high $n_s$ neither approach can explain
$n_s$-independence of the normalized magnetoresistance.

We are grateful to S.V.~Kravchenko for valuable discussions. This
work was supported in part by the Russian Foundation for Basic
Research under Grants No.~00-02-17294 and No.~98-02-16632, the
Programmes "Nanostructures" under Grant No.~97-1024 and "Statistical
Physics" from the Russian Ministry of Sciences, and INTAS under Grant
No.~97-31980.





\end{multicols}

\begin{references}
\bibitem{rem} In a set of experimental publications the $B=0$
metal-insulator transition was detected by a sign change of the
derivative ${\rm d}\rho/{\rm d}T(n_s)$. Alternative methods for
determining the transition point are vanishing activation energy and
vanishing nonlinearity of current-voltage curves as extrapolated from
the insulating phase \cite{shash1,shash}. For the case of Si MOSFETs
all of the definitions are equivalent \cite{com}.
\bibitem{note} We note that in quantizing magnetic fields the values
of $g\approx 5.2$ have been reported at filling factor $\nu=1$
\cite{gap}.
\end{references}
\end{document}